\documentclass[aps, prx, reprint, showpacs, superscriptaddress]{revtex4-1}

\usepackage{mathtools}

\usepackage{paralist}
\usepackage{verbatim}
\usepackage{afterpage}

\usepackage{microtype}
\usepackage{fouriernc}
\usepackage[usenames, dvipsnames, svgnames, table]{xcolor}
\usepackage[colorlinks=true, citecolor=RoyalBlue, 
            linkcolor=BrickRed, urlcolor=ForestGreen]{hyperref}

\usepackage{graphicx}
\usepackage[format=plain, justification=centerlast, small]{caption}
\usepackage{subcaption}

\newcommand{\silica}{SiO$_2$}
\newcommand{\tantala}{Ta$_2$O$_5$}
\newcommand{\algaax}{Al$_x$Ga$_{1-x}$As}
\newcommand{\algaaf}{Al$_{0.92}$Ga$_{0.08}$As}
\newcommand{\checkme}[1]{{\color{orange} #1}}

\newcommand{\mss}[1]{^{(\text{#1})}}
\newcommand{\lowc}{\lowercase}
\newcommand{\rme}{\mathrm{e}}
\newcommand{\rmi}{\mathrm{i}}
\DeclareMathOperator{\real}{Re}

\newcommand{\LigoCaltech}{LIGO Laboratory, California Institute of Technology, MS 100--36, Pasadena, CA 91125, USA}
\newcommand{\NistMD}{Joint Quantum Institute, National Institute of Standards and Technology and University of Maryland, 100 Bureau Drive, Gaithersburg, MD 20899, USA}
\newcommand{\CmsCA}{Crystalline Mirror Solutions LLC, 114 E Haley, Suite N, Santa Barbara, CA 93101, USA}
\newcommand{\CmsWien}{Crystalline Mirror Solutions GmbH, Seestadtstra{\ss}e 27, Top 1.05, A--1220 Vienna, Austria}
\newcommand{\VCQ}{Vienna Center for Quantum Science and Technology (VCQ), Faculty of Physics, University of Vienna, A--1090 Vienna, Austria}
\newcommand{\Fullerton}{Department of Physics, California State University Fullerton, Fullerton, CA 92831, USA}

\begin{document}

\title{Coherent Cancellation of Photothermal Noise in GaAs/Al$_{0.92}$Ga$_{0.08}$As Bragg Mirrors}

\author{Tara Chalermsongsak}
\affiliation{\LigoCaltech}
\author{Evan D. Hall}
\email[Corresponding author: ]{ehall@caltech.edu}
\affiliation{\LigoCaltech}
\author{Garrett D. Cole}
\affiliation{\CmsCA}
\affiliation{\CmsWien}
\author{David Follman}
\affiliation{\CmsCA}
\author{Frank Seifert}
\affiliation{\NistMD}
\author{Koji Arai}
\author{Eric K. Gustafson}
\affiliation{\LigoCaltech}
\author{Joshua R. Smith}
\affiliation{\Fullerton}
\author{Markus Aspelmeyer}
\affiliation{\VCQ}
\author{Rana X Adhikari}
\affiliation{\LigoCaltech}

\begin{abstract}
Thermal noise is a limiting factor in many high-precision optical experiments.
A search is underway for novel optical materials with reduced thermal noise.
One such pair of materials, gallium arsenide and aluminum-alloyed gallium arsenide 
(collectively referred to as AlGaAs), shows promise for its low Brownian noise when 
compared to conventional materials such as silica and tantala.
However, AlGaAs has the potential to produce a high level of thermo-optic noise.
We have fabricated a set of AlGaAs crystalline coatings, transferred to fused silica substrates, 
whose layer structure has been optimized to reduce thermo-optic noise by inducing 
coherent cancellation of the thermoelastic and thermorefractive effects.
By measuring the photothermal transfer function of these mirrors, we find evidence 
that this optimization has been successful.
\end{abstract}

\maketitle

\section{Introduction}
\label{sec:intro}

Many precision optical experiments, such as Advanced LIGO~\cite{Fritschel2014}, employ 
free-space Fabry--P\'{e}rot optical cavities formed from high-performance Bragg mirrors.
Much effort has been put into improving the frequency stability of these optical cavities.
This has involved work on laser frequency stabilization, laser power stabilization (in the 
audio band and at radio frequencies \cite{Zhang2014}), vibration 
isolation~\cite{Chen2006, Ludlow2007, Alnis2008}, and temperature 
stabilization~\cite{Alnis2008}.
Within the last fifteen years, this work has led to optical cavities whose frequency stability is 
limited by thermal noise in the high-reflectivity mirror coatings.
In the case of Advanced LIGO, these coatings are multilayer stacks of ion-beam-sputtered 
silica (\silica) and titania-doped tantala (Ti:\tantala)~\cite{Harry2010}. Experimental 
investigation has revealed that thermal noise in {\silica/\tantala} coatings is dominated 
by mechanical loss in the tantala layers~\cite{Penn2003}. This loss is usually a few 
parts in $10^4$ for undoped coatings, and the addition of titania dopant can reduce this 
loss by at most a factor of two~\cite{Harry2007}.

In the quest for ever more stable Fabry--P\'{e}rot cavities, several groups have looked for 
ways of reducing thermal noise below the limit set by quarter-wavelength (QWL) silica/tantala stacks~\cite{Villar2010, Yam2015, Steinlechner2015a}.
In particular, Cole~et~al.~\cite{Cole2013} fabricated Bragg mirrors from single-crystal 
QWL stacks of gallium arsenide (GaAs) and aluminum-alloyed gallium arsenide (\algaax) 
via a substrate transfer and direct bonding technique.
With these mirrors, they formed an optical cavity whose frequency noise is below the silica/tantala thermal noise level in the band from 1 to 10 Hz, and is consistent with a loss angle below $4\times10^{-5}$, an order of magnitude lower than the dissipation in silica/tantala Bragg mirrors.
Independent ringdown measurements on micromechanical AlGaAs resonators by Cole~et~al.~\cite{Cole2010, Cole2011} confirm this, with measured loss angles as low as $2.4\times10^{-5}$ at room temperature~\cite{Cole2012}.

AlGaAs has been used for nearly 40\,years in optical interference 
coatings~\cite{vanderZiel1975, Jewell1987}, and so its mechanical and optical properties have 
been well characterized.
This makes AlGaAs an attractive coating candidate, since its thermal noise performance 
at room temperature can be accurately modeled.

While AlGaAs is expected to produce lower Brownian noise than silica/tantala (owing to 
its lower mechanical dissipation), it has the potential to produce greater thermo-optic 
noise because the values of the linear coefficient of thermal expansion 
(CTE) $\alpha=(1/L)\partial L/\partial T$ and coefficient of thermorefraction 
(CTR) $\beta = \partial n/\partial T$ are higher in AlGaAs than in silica/tantala.
Coherent cancellation of thermal noise has been described~\cite{Gorodetsky2008, Evans2008, Kimble2008}
for both Brownian noise and thermo-optic noise (driven by temperature fluctuations)
by carefully optimizing the coating's layer structure.
We have carried out such an optimization, with the goal of not only minimizing 
thermo-optic noise, but also controlling the amplitude transmissivity and reflection phase.

\section{Theory of thermal noise}
\label{sec:Theory}

For a Fabry--P\'{e}rot cavity whose mirror surfaces are interrogated by a laser beam, we consider thermal noise terms arising from either mechanical loss or thermal dissipation.
These mechanisms lead to Brownian noise and thermo-optic noise, respectively.
We also consider the effect in which power from the laser beam is absorbed into the mirror, which gives rise to photothermal noise.

\subsection{Brownian noise}
\label{subsec:Brownian}

For a particular material, mechanical loss is quantified by the material's loss angles, which appear as small imaginary components in the material's elastic moduli.
In the case of an isotropic material
\footnote{for cubic single crystals, there are three complex elastic constants to consider}, 
the bulk and shear moduli can be written 
as $K = K_0(1+i\phi_\text{B})$ and $\mu = \mu_0(1+i\phi_\text{S})$, respectively, 
where $\phi_\text{B}$ is the bulk loss angle and $\phi_\text{S}$ is the shear loss 
angle~\cite{Hong2013}.
When the material is subjected to a sinusoidally varying force $F(t) = F_0\cos(2\pi f t)$, these 
loss angles lead to a time-averaged power dissipation 
$W(f) = 2\pi f(U_\text{B} \phi_\text{B} + U_\text{S} \phi_\text{S})$, where $U_\text{B}$ is the 
maximum energy stored in bulk deformation, and $U_\text{S}$ is the maximum energy 
stored in shear deformation.

According to the fluctuation--dissipation theorem (FDT), mechanical dissipation leads to 
fluctuation in the generalized coordinate $x$ conjugate to $F$~\cite{Callen1951}.
Using the ``direct approach''~\cite{Gonzalez1994, Levin1998}, the power spectral density 
(PSD) of the fluctuations in $x$ is given by
    \begin{equation}
        S_x(f) = \frac{2k_\text{B} T}{\pi^2 f^2} \frac{W(f)}{F_0^{\,2}}.
        \label{eq:FDT}
    \end{equation}
With $W$ computed as above, $S_x(f)$ is the Brownian noise in the material.

\subsection{Thermo-optic noise}
\label{subsec:ThermoOptic}

Thermo-optic noise is a consequence of thermodynamically driven temperature fluctuations 
within a material.
Again using the fluctuation--dissipation theorem, the PSD $S_T(f)$ of temperature fluctuations 
can be computed by considering the generalized force conjugate to temperature; namely, the entropy~\cite{Levin2008}.
With $S_T(f)$ in hand, the thermo-optic contribution to the cavity length noise $S_L(f)$ can be 
computed via the CTE $\alpha = (1/L)\partial L/\partial T$ (giving thermoelastic noise, with 
PSD $S_x\mss{TE}$) and the CTR $\beta = \partial n/\partial T$ (giving thermorefractive 
noise, with PSD $S_x\mss{TR}$)~\cite{Evans2008, Ballmer2015}.

\subsection{Photothermal noise}

The coefficients of thermal expansion and thermorefraction can manifest themselves not only through thermo-optic noise, but also through photothermal noise, in which laser power is absorbed in the coating and subsequently produces fluctuations in the coating's temperature.
In the frequency band of interest, the thermodynamic fluctuations and the laser power fluctuations have thermal lengths which are much larger than the coating thickness.  
Therefore, the coherent cancellation of the thermodynamically induced thermoelastic and thermorefractive noises will also occur for the noises induced by laser power fluctuation (i.e., the photothermal noise).
Since the level of photothermal noise is proportional to the absorbed power, its effect can be enhanced for observation by modulating the power incident on the mirrors. 
By observing the cancellation of photothermal noise, we can use it as evidence for the cancellation of thermo-optic noise. 
The details on the measurement will be discussed in section \ref{sec:Experiment}.

\section{Noise budget for Fabry--P\'{e}rot cavities with A\lowc{l}G\lowc{a}A\lowc{s}-coated mirrors}
\label{sec:NoiseBudget}

\begin{table*}[tbp]
    \centering
    \begin{tabular}{l l c p{2.7in}}
        \toprule
        Symb.               & Description                    & Value    & Notes \\
        \colrule
        $L$                 & Nominal spacer length          & 36.8(3)\,mm
            &   \\
        $R_\text{sp}$       & Outer spacer radius            & 19.0\,mm
            &   \\
        $r_\text{sp}$       & Inner spacer radius            & 5.1\,mm
            &   \\
        $R_\text{s}$        & Mirror substrate radius        & 25.4\,mm
            &   \\
        $\mathcal{R}$       & Mirror radius of curvature     & 1000(5)\,mm
            & As specified before application of coatings.
              0.5\,\% uncertainty assumed. \\
        $\lambda_0$         & Laser vacuum wavelength        & 1064\,nm
            &   \\
        $w$                 & Spot size on mirrors           & 215.4(5)\,{\textmu}m
            & Defined as the radius for which the beam intensity $I$ satisfies $I(w) = I(0)/\rme^2$.
              Computed as $w = (\lambda_0\mathcal{R}/\pi)^{1/2}/(2\mathcal{R}/L-1)^{1/4}$.  \\
        $\mathcal{F}$       & Finesse                        & 16700(1400), 17600(1600)
            & From measurement of the cavity poles.  \\
        $\mathcal{T}$       & Transmissivity (per mirror)    & 153(7)\,ppm
            & Average over five mirrors fabricated for this work.  \\
        $\mathcal{A}$       & Absorption (per mirror)        & 4.96(23)\,ppm, 5.19(27)\,ppm
            & From measurement of the photothermal transfer function.  \\
        $\mathcal{S}$       & Scatter (per mirror)           & 30(17)\,ppm, 20(18)\,ppm
            &  Calculated as $\mathcal{S} = \pi/\mathcal{F} - \mathcal{T} - \mathcal{A}$ \\
        $T$                 & Cavity temperature             & 305(1)\,K
            &   \\
        \colrule
        $E_\text{s}$        & Young's modulus of fused silica        & 72(1)\,GPa
            &   \\
        $\sigma_\text{s}$   & Poisson's ratio of fused silica        & 0.170(5)
            &   \\
        $\kappa_\text{s}$   & Thermal conductivity of fused silica & 1.38\,W/(m\,K)
            &   \\
        $C_\text{s}$        & Heat capacity of fused silica        & $1.6\times10^6$\,J/(m$^3$\,K)
            &   \\
        $\alpha_\text{s}$   & CTE of fused silica                  & $5.1\times10^{-7}$\,K$^{-1}$
            &   \\ 
        $n_\text{s}$        & Refractive index of fused silica     & 1.46
            &   \\
        $\phi_\text{s}$     & Loss angle of fused silica           & $1\times10^{-7}$
            &   \\
        \colrule
        $x$                  & Aluminum alloying fraction    & 0.920(6) & \\
        $E_\text{L}$         & Young's modulus of \algaaf      & 100(20)\,GPa
            & Nominal value from Cole~et~al.~\cite{Cole2013}.
              20\,\% uncertainty assumed. \\
        $E_\text{H}$         & Young's modulus of GaAs         & 100(20)\,GPa
            & See note for $E_\text{L}$.   \\
        $\sigma_\text{L}$    & Poisson's ratio of \algaaf      & 0.32(3)
        & After Adachi~\cite[p.~24]{Adachi1993}.
              10\,\% uncertainty assumed. \\
        $\sigma_\text{H}$    & Poisson's ratio of GaAs         & 0.32(3)
            & See note for $\sigma_\text{L}$. \\
        $\kappa_\text{L}$    & Thermal conductivity of \algaaf   & 70(4)\,W/(m\,K)
            & Computed as $(55 - 212x + 248x^2) \text{ W m}^{-1} \text{ K}^{-1}$, from the Ioffe Institute~\cite{IoffeAlGaAs}.
              5\,\% uncertainty assumed. \\
        $\kappa_\text{H}$    & Thermal conductivity of GaAs      & 55(3)\,W/(m\,K)
            & See note for $\kappa_\text{L}$.   \\
        $C_\text{L}$         & Heat capacity of \algaaf      & $1.70(9)\times10^6$\,J/(m$^3$\,K)
            & Computed as $(1.75 + 0.11x - 0.19x^2)\times10^6~\text{J/(m$^3$\,K)}$, from Adachi~\cite[p.~41]{Adachi1993} and the Ioffe Institute~\cite{IoffeAlGaAs}.\footnote{From the Ioffe Institute, the density of {\algaaf} is $\rho = (5.32 - 1.56x)\times10^{3}\text{ kg m}^{-3}$.
            From Adachi, the specific heat capacity is $\mathcal{C} = (320 + 132x) \text{ J/(kg K)}$.
            The volumetric heat capacity is then $C = \mathcal{C}\rho$.}
            5\,\% uncertainty assumed. \\
        $C_\text{H}$         & Heat capacity of GaAs         & $1.75(9)\times10^6$\,J/(m$^3$\,K)
            & See note for $C_\text{L}$.   \\
        $\alpha_\text{L}$    & CTE of \algaaf                & $5.2(3)\times10^{-6}\,\text{K}^{-1}$
        & Computed as $(5.73 - 0.53x)\times10^{-6}~\text{K}^{-1}$, after Levinshtein et al~\cite[p.\,4]{Levinshtein1996}.
              5\,\% uncertainty assumed. \\
        $\alpha_\text{H}$    & CTE of GaAs                   & $5.7(3)\times10^{-6}\,\text{K}^{-1}$
            & See note for $\alpha_\text{L}$.   \\
        $\beta_\text{L}$     & CTR of \algaaf                & $179(7)\times10^{-6}\,\text{K}^{-1}$
            & Computed as $[366(7) - 203x]\times10^{-6}~\text{K}^{-1}$, after measurements on GaAs and AlAs by Talghader and Smith~\cite{Talghader1995,Talghader1996}. \\
        $\beta_\text{H}$     & CTR of GaAs                   & $366(7)\times10^{-6}\,\text{K}^{-1}$
            & See note for $\beta_\text{L}$.   \\
        $n_\text{L}$         & Refractive index of \algaaf   & 2.98(3)
            & After Cole~et~al.~\cite{Cole2013}.
              1\,\% uncertainty assumed. \\
        $n_\text{H}$         & Refractive index of GaAs      & 3.48(3)
            & See note for $n_\text{L}$.   \\
        \colrule
        $\alpha_\text{c}$    & Effective CTE of coating      & $19(3)\times10^{-6}\,\text{K}^{-1}$
            & Computed in appendix \ref{sec:TOparams}   \\
        $\beta_\text{c}$     & Effective CTR of coating      & $79(4)\times10^{-6}\,\text{K}^{-1}$
            & Computed in appendix \ref{sec:TOparams}   \\
        $C_\text{c}$         & Effective heat capacity of coating   & $1.73(6)\times10^6\,\text{J/(m$^3$\,K)}$
            & Computed in appendix \ref{sec:TOparams}   \\
        $\kappa_\text{c}$    & Effective thermal conductivity of coating   & $61.6(2.4)\,\text{W/(m\,K)}$
            & Computed in appendix \ref{sec:TOparams}   \\
        $N$                  & Number of coating layers      & 57
            &   \\
        $d$                  & Total thickness of coating    & 4.6806(4)\,{\textmu}m
            & Physical thickness, assuming 50\,pm random uncertainty in each layer.
        There is additional systematic uncertainty; see appx.~\ref{sec:CoatingStructure}. \\
        $\phi_\text{c}$      & Effective coating loss angle  & $2.41(20)\times10^{-5}$
            & From ringdown measurements by Cole et~al. on QWL AlGaAs microresonators~\cite{Cole2011}. \\
        \botrule
    \end{tabular}
    \caption{Parameters for our reference cavities.}
    \label{tab:CavityParams}
\end{table*}

The optical, mechanical, and material parameters for our reference cavities are given in 
table~\ref{tab:CavityParams}. Where two values are given, these refer to the values measured 
for each of the two cavities.

In this section we give a concise overview of the noise budget for our fixed-spacer 
Fabry--P\'{e}rot cavities.
A more detailed explanation of such noise budgeting has already been given by 
Chalermsongsak~et~al.~\cite{Chalermsongsak2014}.

The length noise $S_L(f)$ of a fixed-spacer cavity is
    \begin{align}
        S_L &= 2S_x\mss{cBr} + 2S_x\mss{cTO} + 2S_x\mss{subBr} + 2S_x\mss{subTE} \nonumber \\
            &\hphantom{=} \hspace{2em} + S_L\mss{spBr} + S_L\mss{spTE} + 4S_x\mss{PT}.
        \label{eq:CavityLengthNoise}
    \end{align}

\subsection{Coating noise}
The quantity $S_x\mss{cBr}$ is the effective displacement of each mirror's position due 
to coating Brownian noise.
By treating the coating as a thin, homogeneous layer of material with a Young's 
modulus $E_\text{c}$ and a Poisson's ratio $\sigma_\text{c}$, one can write this 
noise as~\cite{Cole2013}
    \begin{align}
        S_x\mss{cBr}(f) &= \frac{4 k_\text{B} T}{\pi^2 f}
            \frac{d}{w^2 E_\text{s}^2} \frac{\phi_\text{c}}{E_\text{c} \bigl(1-\sigma_\text{c}^2\bigr)} \nonumber\\
            &\hphantom{=} \quad\times
            \left[E_\text{c}^2 (1+\sigma_\text{s})^2(1-2\sigma_\text{s})^2
                + E_\text{s}^2(1+\sigma_\text{c})^2(1-2\sigma_\text{c})\right],
        \label{eq:CoatingBrownian}
    \end{align}
where $\phi_\text{c}$ is the coating's loss angle.
This loss angle is a linear combination of the bulk and shear loss angles of the individual
coating materials.

The quantity $S_x\mss{cTO}$ is the effective displacement of each mirror's position due
to coating thermo-optic noise~\cite{Evans2008}:
    \begin{equation}
        S_x\mss{cTO}(f) = S_T(f) \ \Gamma_\text{tc}(f) \,
            \left(\alpha_\text{c} d - \beta_\text{c} \lambda_0
            - \alpha_\text{s} d C_\text{c} / C_\text{s}\right)^2,
        \label{eq:CoatingThermoOptic}
    \end{equation}
with~\cite[eqs.~3.27--8]{Martin2013}
    \begin{equation}
        S_T(f) = \frac{2^{3/2} k_\text{B} T^2}{\pi\kappa_\text{s} w} M\bigl(f/f_\text{T}\bigr),
        \label{eq:tempPSD}
    \end{equation}
where $f_\text{T} = \kappa_\text{c} / (\pi w^2 C_\text{c})$ and
    \begin{equation}
        M(\Omega) = \int\limits_0^\infty\! \mathrm{d}u\;\real\left[\frac{u\,\rme^{-u^2/2}}{\left(u^2-\rmi \Omega\right)^{1/2}}\right].
    \end{equation}
$\Gamma_\text{tc}$ is a correction factor accounting for the nonzero thickness of the coating~\cite[eq.~39]{Evans2008}.

    \begin{figure}[tbp]
        \centering
        \begin{subfigure}[b]{0.5\textwidth}
            \includegraphics[width=\textwidth]{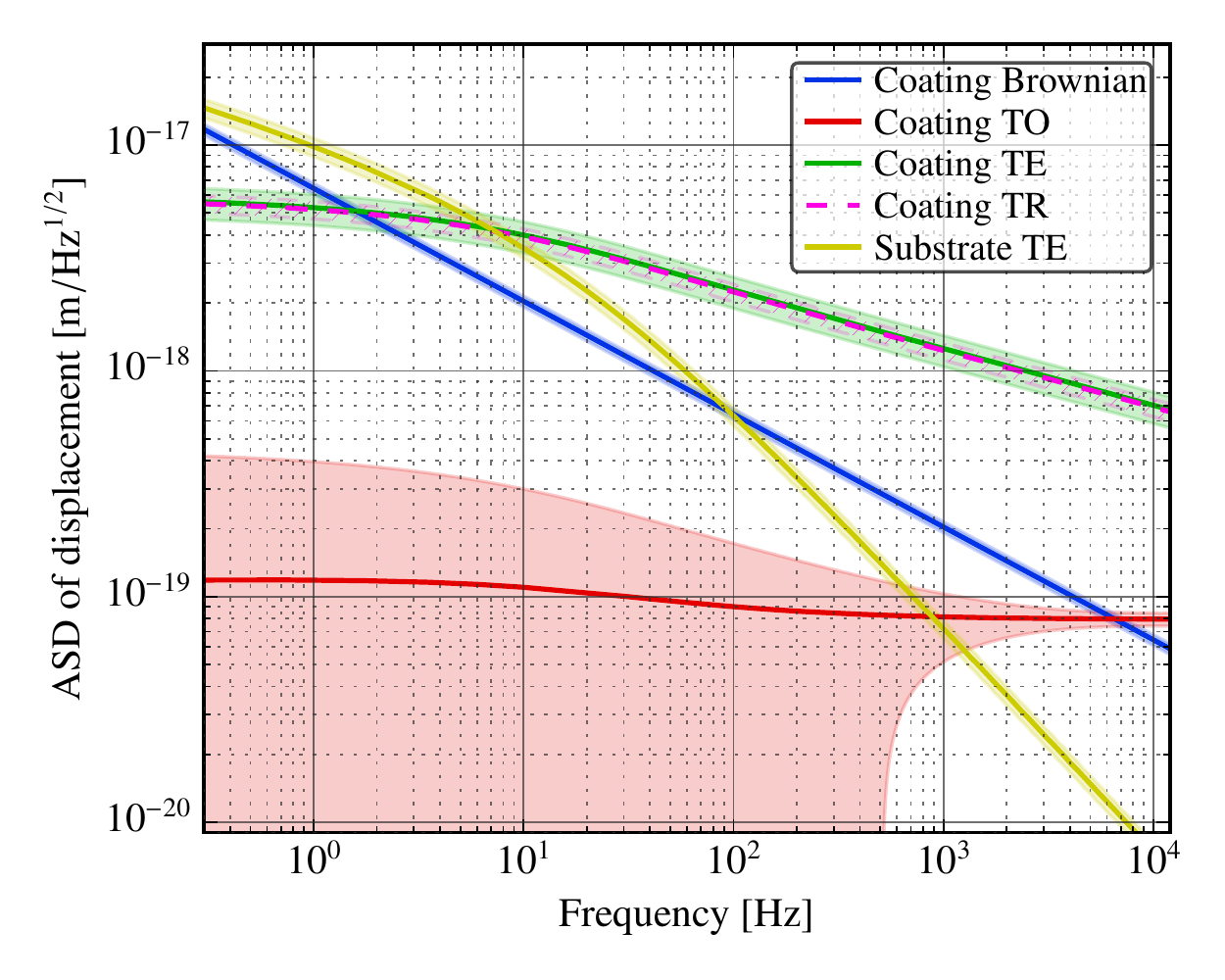}
            \caption{Anticipated thermal noise sources for our optimized AlGaAs mirrors.}
            \label{fig:ThermoOpticPlotOpt}
        \end{subfigure}

        \begin{subfigure}[b]{0.5\textwidth}
            \includegraphics[width=\textwidth]{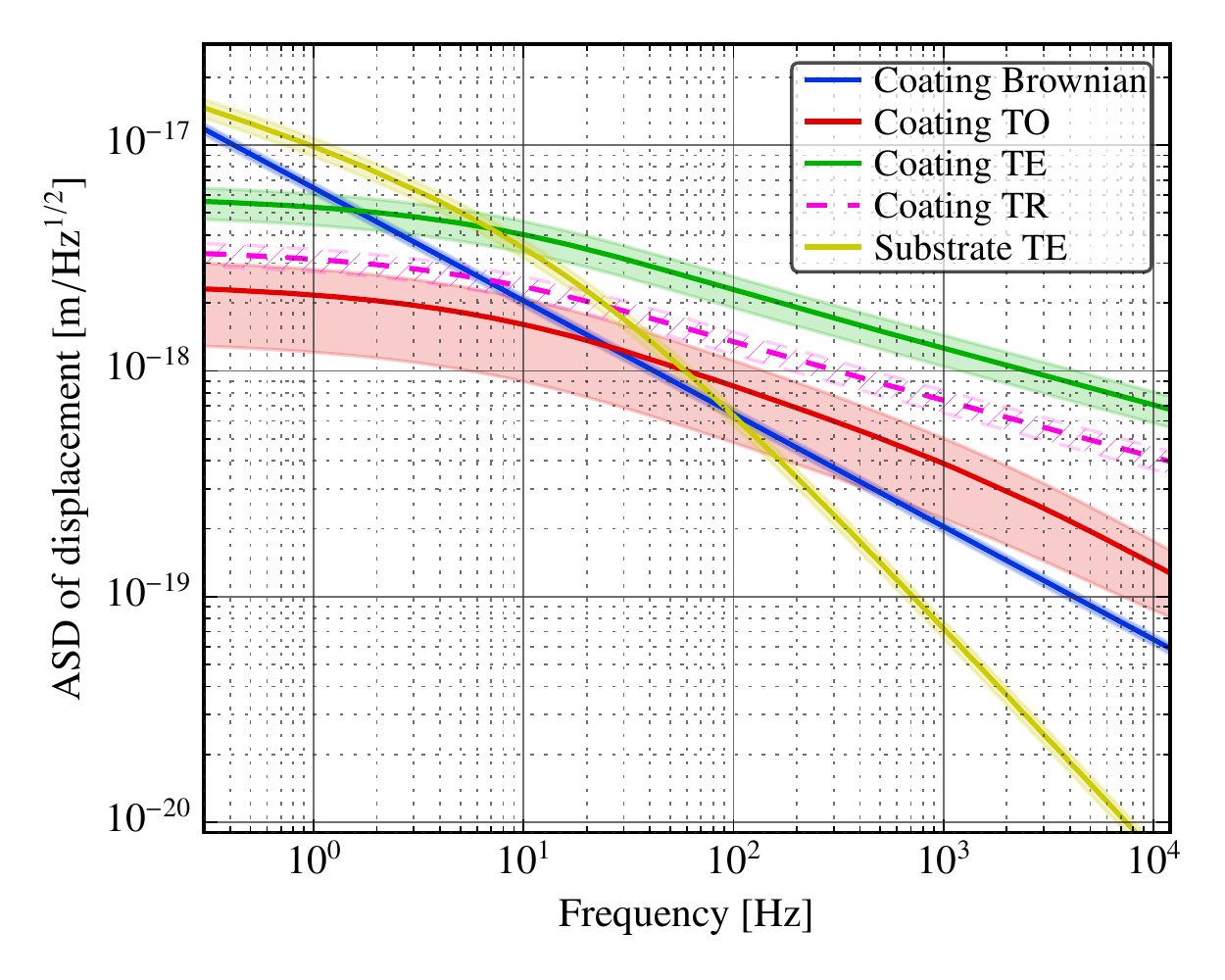}
            \caption{Thermal noise sources for a hypothetical 57-layer QWL coating, with the first and last layers being GaAs.}
            \label{fig:ThermoOpticPlotQWL}
        \end{subfigure}
        \caption{Important thermal noise sources for our optimized AlGaAs mirrors (fig. \ref{fig:ThermoOpticPlotOpt}) and for a 57-layer QWL stack (fig. \ref{fig:ThermoOpticPlotQWL}).
            The shaded/hatched regions indicate $1\sigma$ uncertainties propagated forward from the uncertainties given in table \ref{tab:CavityParams}.
            Our optimized coatings are designed to maximize the cancellation between the TE and TR noise terms, resulting in TO noise that is subdominant to the coating Brownian and substrate thermoelastic noises below 5\,kHz.}
        \label{fig:ThermoOpticPlot}
    \end{figure}
The optimized structure of our coatings results in thermo-optic noise whose PSD 
$S_x\mss{cTO}$ is lower than the incoherent sum $S_x\mss{cTE} + S_x\mss{cTR}$ of the 
thermoelastic and thermorefractive noise terms (the first two terms of 
Eq.~\ref{eq:CoatingThermoOptic}, respectively).
In figure~\ref{fig:ThermoOpticPlot} we plot the anticipated thermo-optic, thermoelastic, 
thermorefractive, and Brownian noises of our coatings, along with the same noises for a 
hypothetical 57-layer QWL coating whose first and last layers are GaAs.

\subsection{Substrate and spacer noise}
By computing the substrate thermoelastic noise $S_x\mss{subTE}$ following 
Cerdonio~et~al.~\cite{Cerdonio2001}, we anticipate that $S_x\mss{subTE}$ will be a 
limiting noise source for our cavities below 100\,Hz.
Additionally, we have analytically computed the substrate Brownian noise $S_x\mss{subBr}$ 
following Levin~\cite[eq.~2]{Levin1998} and Liu~et~al.~\cite[eq.~59]{Liu2000}, and have 
found that this noise is much less than the coating Brownian noise at all frequencies of interest.
In the case of the spacers, we used a finite-element model to compute the Brownian and 
thermo-elastic noises, and have found them to be negligible as well.
For details, we again refer the reader to the full discussion in 
Chalermsongsak~et~al.~\cite{Chalermsongsak2014}.
In the rest of this work, we only consider the thermal noise mechanisms arising from the 
coatings, and from the thermoelastic noise of the fused silica substrates.

\subsection{Photothermal noise}
 
The quantity $S_x\mss{PT}$ is the effective displacement of each mirror's position due to 
photothermal noise: 
    \begin{equation}
        S_x\mss{PT}(f) = \bigl|H\mss{PT}(f)\bigr|^2 S_P(f),
        \label{eq:Photothermal}
    \end{equation}
where $H\mss{PT}(f)$ is the photothermal transfer function which takes intracavity 
power fluctuation $P(f) = P_0 / (1+\rmi f/f_\text{cav})$ to displacement $x(f)$.
It has coating thermoelastic, coating thermorefractive, and substrate thermoelastic 
contributions which are described in detail by Farsi~et~al.~\cite[appx.~A]{Farsi2012}:
    \begin{equation}
        H\mss{PT}(f) = H\mss{cTE}(f) + H\mss{cTR}(f) + H\mss{sTE}(f).
        \label{eq:PhotothermalTF}
    \end{equation}
The noise $S_x\mss{PT}(f)$ enters into equation \ref{eq:CavityLengthNoise} with a factor 
of 4 because the photothermal fluctuations in the two mirrors are driven coherently by 
the field circulating in the cavity.

By applying the formalism of Farsi et al. to our coating design \footnote{In the formalism 
of Farsi~et~al., the relevant coating CTE is a simple volumetric average 
$\alpha_\text{c}' =  \sum_j \alpha_j d_j / d$ rather than the CTE $\alpha_\text{c}$ 
presented in appendix~\ref{sec:TOparams} of this work.}, we anticipate that our 
coatings should display coherent cancellation of $H\mss{cTE}$ and $H\mss{cTR}$.
In figure~\ref{fig:PhotothermalMeasurement} we plot the expected photothermal transfer 
function of our coatings.

\section{Optimization of coating structure}
\label{sec:Optimization}

\begin{figure}[tbp]
    \centering
    \includegraphics[width=0.49\textwidth]{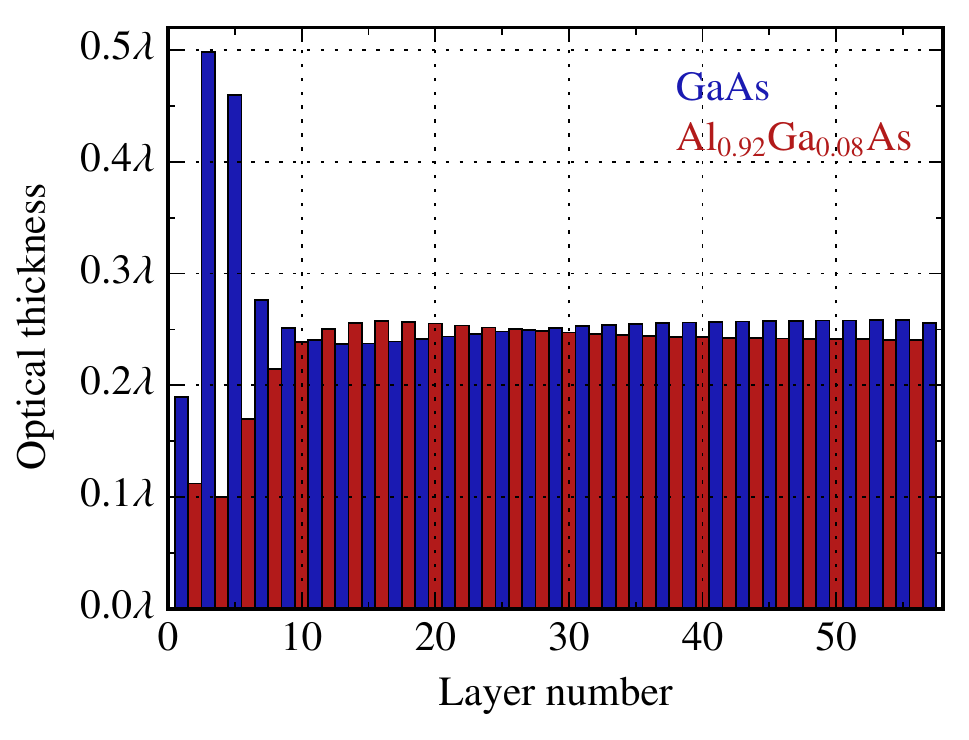}
    \caption{Optimized layer structure of our AlGaAs coatings.
    Each layer is expressed in optical thickness, so that $\lambda = \lambda_0/n_\text{L}$ 
    for the low-index (\algaaf) layers and $\lambda = \lambda_0/n_\text{H}$ for the 
    high-index (GaAs) layers.}
    \label{fig:AlgaasOptThick}
\end{figure}

Examining \eqref{eq:CoatingThermoOptic}, it is evident that the thermo-optic noise can 
be nearly cancelled by manufacturing a mirror whose material parameters 
satisfy $(\alpha_\text{c} - \alpha_\text{s} C_\text{c} / C_\text{s}) d = \beta_\text{c} \lambda_0$.

In designing our AlGaAs Bragg mirrors, we wanted a layer structure that would produce 
a coating with
    \begin{inparaenum}[(1)]
        \item minimal thermo-optic noise $S_x\mss{cTO}$ in our band of interest (10\,Hz to 1\,kHz),
        \item a transmissivity $\mathcal{T}$ close to our target $\mathcal{T}_0 = 200$\,ppm, and
        \item a reflected phase $\Delta$ that is close to $\Delta_0 = 180^\circ$~\footnote{The 
            condition $\Delta_0 = 180^\circ$ ensures that the field at the surface of the coating is
            nearly zero. This reduces the likelihood that contaminants such as dust will be
            burned onto the coating~\cite[\S 5.2.2]{Chalermsongsak2014b}.}.
    \end{inparaenum}
To that end, we constructed an initial cost function
\begin{align}
    y\bigl[S_x\mss{cTO}(f_0), \mathcal{T}, \Delta\bigr] &= w_1 \, S_x\mss{cTO}(f_0)
        + w_2 \left(\frac{\mathcal{T} - \mathcal{T}_0}{\mathcal{T}_0}\right)^2 \nonumber \\
        &\hphantom{=} \hspace{2em} + w_3 (\Delta - \Delta_0)^2,
    \label{eq:AlgaasCostFunc}
\end{align}
where $w_1$, $w_2$, and $w_3$ are weights, and $f_0 = 100$\,Hz.
The quantities $S_x\mss{cTO}(f_0)$, $\mathcal{T}$, and $\Delta$ are functions of the layer structure $\mathbf{d} = \begin{pmatrix}d_1 & d_2 & d_3 & \cdots & d_{57}\end{pmatrix}^T$.

Then, to ensure that our optimization was robust against small uncertainties in the indices of refraction $n_\text{L}$ and $n_\text{H}$, we constructed a modified cost function:

\begin{equation}
    y' = \sum_{n_\text{H}}\sum_{n_\text{L}} y\bigl[S_x\mss{cTO}(f_0), \mathcal{T}, \Delta; n_\text{H}, n_\text{L}\bigr]
    \label{eq:AlgaasModCostFunc}
\end{equation}
with $n_\text{H} \in \{3.47, 3.50, 3.53\}$ and $n_\text{L} \in \{2.97, 3.00, 3.03\}$.
We then numerically minimized this cost function in order to find an optimal layer structure $\mathbf{d}$.

\begin{figure}[!tbp]
    \centering
    \includegraphics[width=0.48\textwidth]{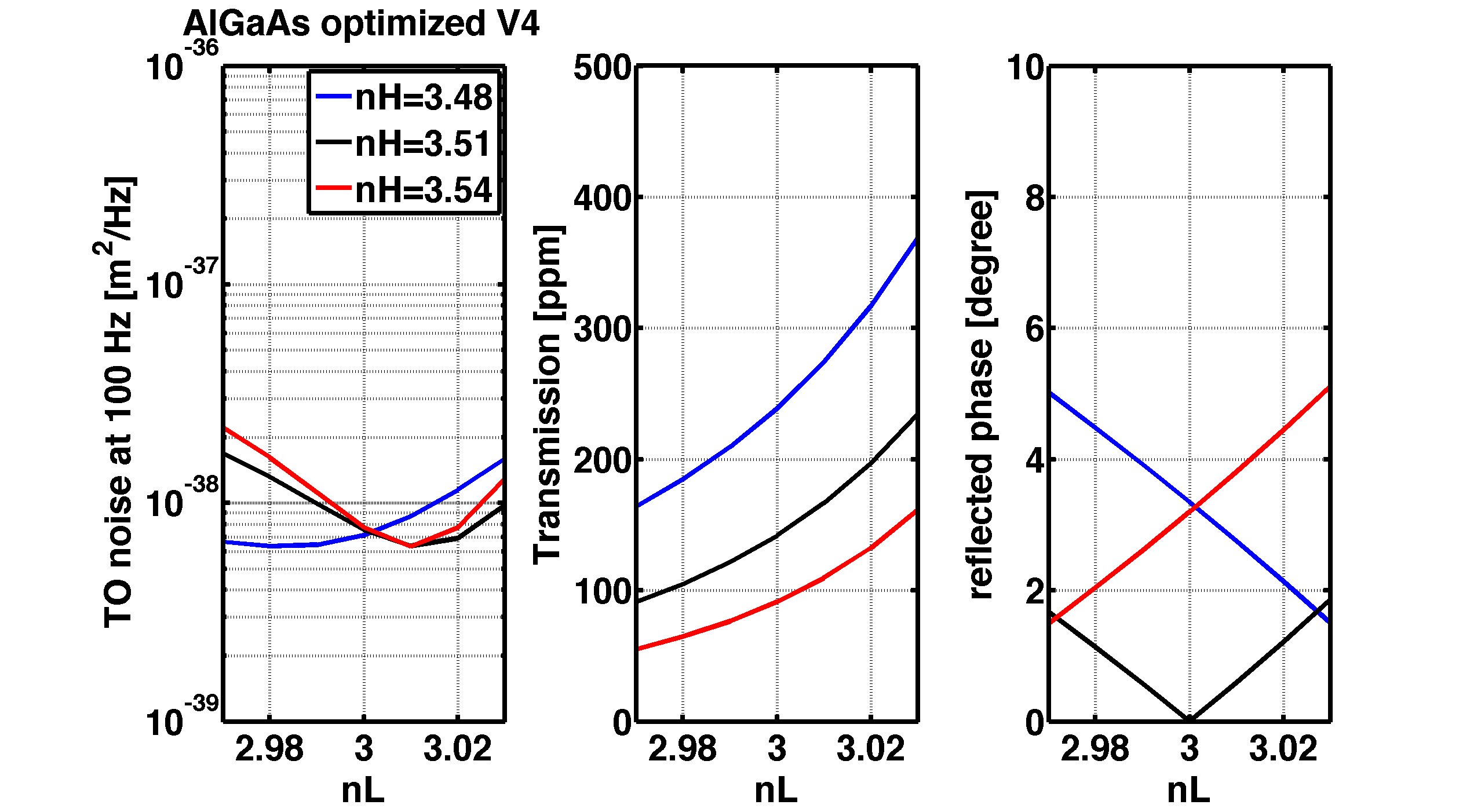}
    \caption{Expected thermo-optic noise (at 100\,Hz), transmissivity, and reflection phase
    (modulo $180^\circ$) 
    of our optimized coating, given small uncertainties in $n_\text{L}$ and $n_\text{H}$.}
    \label{fig:OptimizationResults}
\end{figure}
To test the robustness of our layer structure against other material parameters, we ran a Monte Carlo simulation using the parameter values and uncertainties for the heat capacities, 
thermal conductivities, Young's moduli, Poisson's ratios, CTEs, and CTRs of 
GaAs and {\algaaf}.
We also took into account possible systematic and statistical errors in the thickness 
of each coating layer.
The simulation shows that 70\,\% of the trials result in coatings that satisfy
    \begin{inparaenum}[(1)]
        \item $\sqrt{S_x\mss{cTO}(f_0)} < 3.9\times10^{-20}\,\text{m/Hz}^{1/2}$,
        \item $100\,\text{ppm} < \mathcal{T} < 300\,\text{ppm}$, and
        \item $|\Delta - 180^\circ| < 7^\circ$.
    \end{inparaenum}

In figure~\ref{fig:AlgaasOptThick}, we plot the structure of our optimized AlGaAs coating.
The numerical values for each coating layer are given in table~\ref{tab:CoatStruct} in the appendix.
In figure~\ref{fig:OptimizationResults} we plot the expected performance of our optimized structure as a function of $n_\text{L}$ and $n_\text{H}$.

\section{Experiment}
\label{sec:Experiment}

\begin{figure*}[!htbp]
    \centering
    \includegraphics[width=0.9\textwidth]{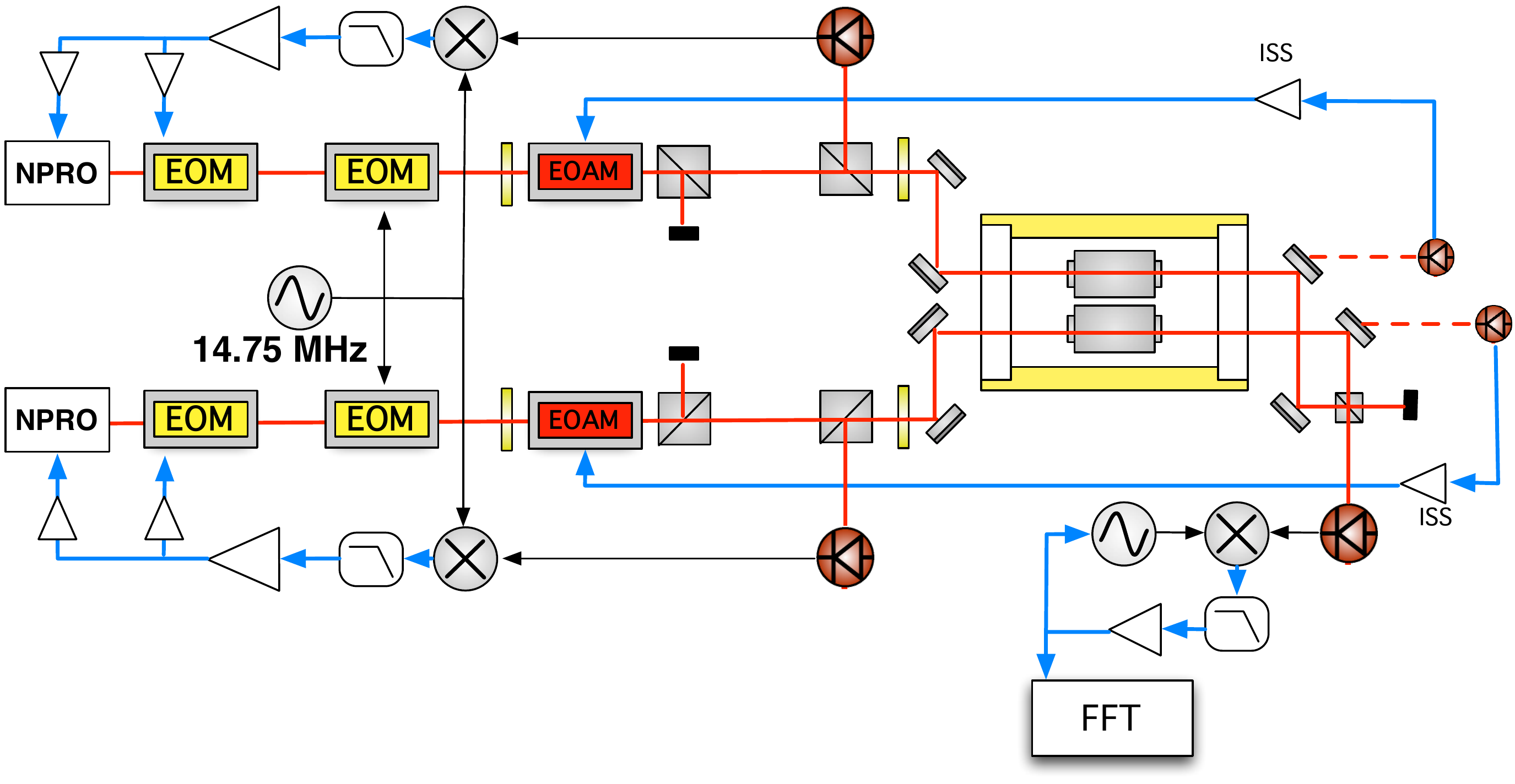}
    \caption{Diagram of experimental setup used to measure the beat note between two 3.68-cm Fabry--P\'{e}rot cavities with AlGaAs mirror coatings.
    The box labeled ``FFT'' is an audio-band signal analyzer which is capable of measuring both PSDs and swept-sine transfer functions.}
    \label{fig:ExperimentalSetup}
\end{figure*}

A diagram of the experiment is shown in figure~\ref{fig:ExperimentalSetup}.
This setup has been described in detail previously~\cite{Chalermsongsak2014}; we briefly 
summarize the salient points here.

We independently lock a 1064\,nm non-planar ring oscillator (NPRO) laser to each cavity.
We use the Pound--Drever--Hall interrogation technique \cite{Drever1983, Hils:1987vb} 
to derive an error signal which indicates the detuning of the laser from cavity resonance.
This error signal is electronically amplified and filtered to produce a control signal.
At frequencies from DC to tens of kilohertz, this control voltage is applied to a 
piezoelectric transducer (PZT) attached to the NPRO crystal.
At frequencies from tens to hundreds of kilohertz, the control voltage is applied to a 
broadband, MgO:LiNbO$_3$ electro-optic modulator (EOM).
This servo loop achieves a unity-gain frequency above 700\,kHz.
Within the bandwidth of the servo, the laser frequency is made to tightly track the length 
fluctuation of the cavity.

The transmitted beams from the two cavities are interfered on an RF photodiode, producing an RF beat note with a frequency $\nu_\text{b} = \nu_2 - \nu_1$ of approximately 10\,MHz.
Using a voltage-controlled oscillator (VCO) of similar frequency, we form a 
high-gain phase-locked loop (PLL). The control signal of this PLL is proportional to the
differential frequency fluctuations of the two cavities.

The experiment is additionally equipped with an intensity stabilization servo (ISS).
For each cavity, a portion of the transmitted beam is directed onto a large-area 
photodiode.
The resulting audio-band signal is amplified and fed back to an electro-optic 
amplitude modulator (EOAM), thereby stabilizing the cavity power in the frequency 
band of interest.

\section{Results}
\label{sec:Results}

\subsection{Photothermal transfer function}

\begin{figure}[!tbp]
    \centering
    \includegraphics[width=\columnwidth]{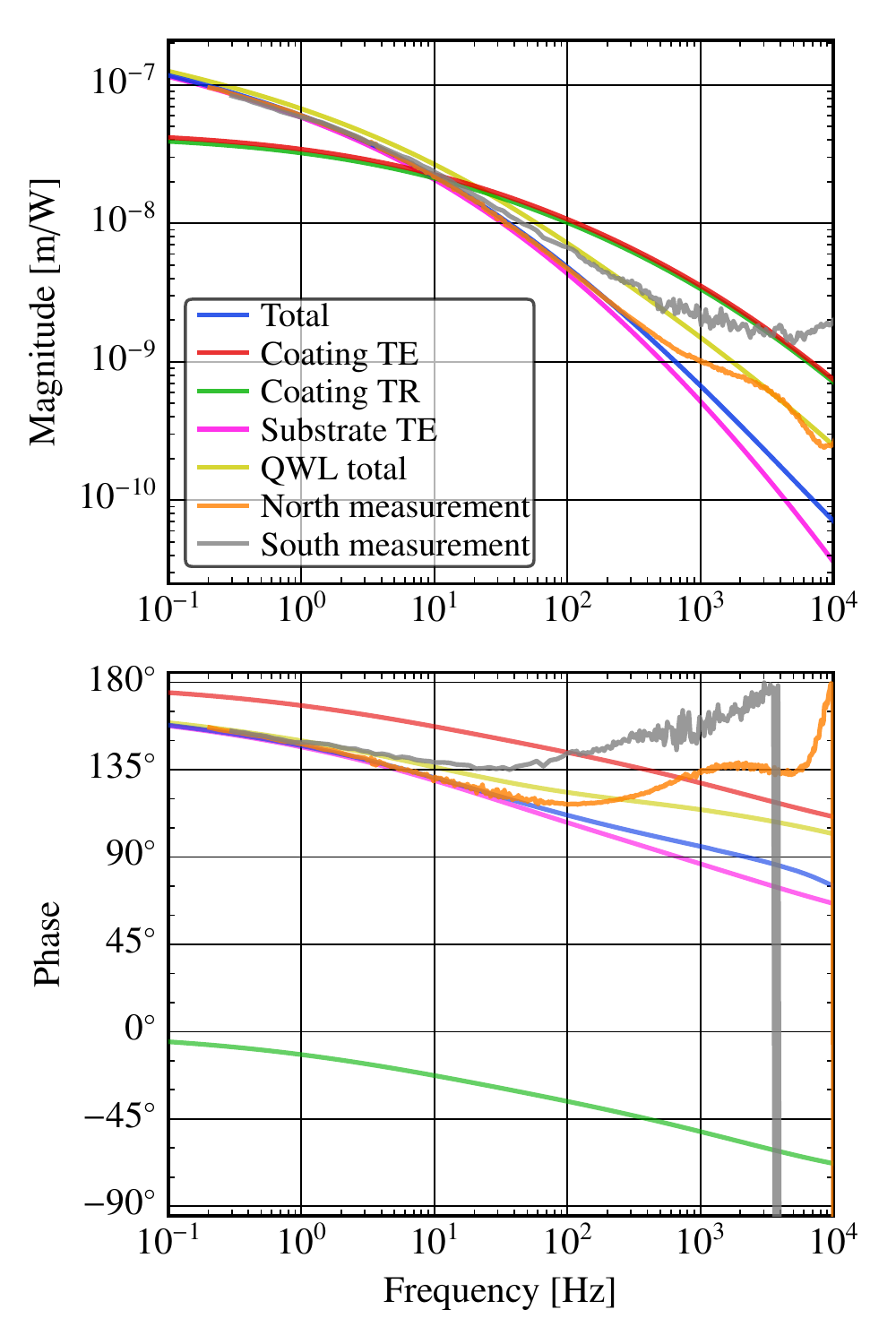}
    \caption{Measured photothermal transfer function, along with expected transfer 
        function after Farsi~et~al.~\cite{Farsi2012}, for our two cavities (labeled ``north''
        and ``south'').
      The units are chosen so that the transfer function takes the absorbed power (per mirror) 
      to the shift in the sensed displacement of each mirror.
      In this plot we also show the total photothermal transfer function that would be 
      expected from a 81-layer QWL coating constructed from GaAs and \algaaf.}
    \label{fig:PhotothermalMeasurement}
\end{figure}

For each cavity, we measured the photothermal transfer function as follows.
With the ISS disengaged, we drove the EOAM with a sinusoidal voltage $V(f)$.
We then measured the transfer function which takes the transmitted cavity power 
$P_\text{trans}(f)$ (measured via a calibrated photodiode) to beat note fluctuation $\nu_\text{b}(f)$.
The measured transfer function $\nu_\text{b}(f)/P_\text{trans}(f)$ can be used to estimate 
the per-coating absorption 
$\mathcal{A}$ by comparing with the expected transfer function $H\mss{PT}(f) = \nu_\text{b}(f)/P_\text{abs}(f)$ in equation~\ref{eq:PhotothermalTF}.
Here $P_\text{abs}(f) = \mathcal{A}\,P_\text{circ}(f) = \mathcal{A}\,P_\text{trans}(f)/\mathcal{T}$, and $\mathcal{T} = 153(7)$\,ppm is the measured per-mirror transmissivity (to be compared to the value of 151\,ppm, computed from the nominal coating design).

To obtain an estimate of $\mathcal{A}$, we write our measurement as $\hat{H}\mss{PT}(f) = \mathcal{T} \nu_\text{b}(f) / \mathcal{A}\, P_\text{trans}(f)$ and then minimize the sum
    \begin{equation}
        \sum_{f_i \le 10\,\text{Hz}} \left|\, {\log_{10}{\frac{\hat{H}\mss{PT}(f_i)}{H\mss{PT}(f_i)}}}\right|^2
    \end{equation}
with respect to $\mathcal{A}$.
For each of the cavities, we find mean per-mirror absorptivities of 4.96(1)\,ppm and 5.19(13)\,ppm, where the uncertainties are statistical.
To these uncertainties we must add the uncertainty in the value of the transmission $\mathcal{T}$ used to estimate the circulating cavity power.
With this uncertainty incorporated, our measured absorptivities are 4.96(23)\,ppm and 5.19(27)\,ppm.
For our coatings, the penetration depth is $\delta = 560\,\text{nm}$, and so the absorption coefficient is $\alpha = \mathcal{A}/(2\delta) \sim 0.05\,\text{cm}^{-1}$.

In figure~\ref{fig:PhotothermalMeasurement} we show the measured photothermal transfer functions for our AlGaAs coatings, along with the expectation.
Below 10~Hz, the magnitude and phase agree well with the expectation, particularly for the north cavity.
From 10~Hz to 3~kHz, the agreement between measurement and expectation is imperfect, and the cause of the discrepancy at high frequencies is not understood; it may be laser power fluctuation coupling into the PLL.
However, it is evident that the measured transfer functions have magnitudes that lie below the individual thermoelastic and thermorefractive contributions to the expected photothermal transfer function.
We take this as strong evidence that thermo-optic cancellation has been successfully realized in these coatings.  
\subsection{Scatter}

\begin{figure}[!tbp]
    \centering
    \includegraphics[width=\columnwidth]{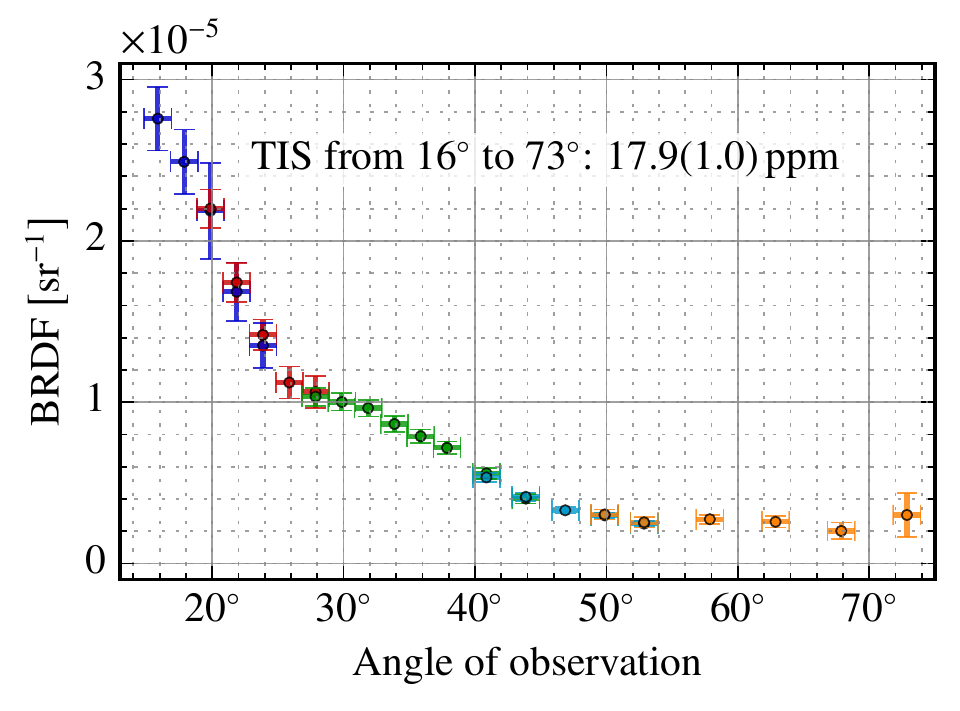}
    \caption{A representative BRDF measurement of our AlGaAs coatings.
    Different colors denote different exposure times for the CCD acquisition.}
    \label{fig:ScatterMeasurement}
\end{figure}

To assess the scatter of our AlGaAs coatings, we used the technique described by Maga\~{n}a-Sandoval et~al.~\cite{Magana2012} to measure the bidirectional reflectance distribution function (BRDF) of our coatings.
With a laser normally incident on the coating, light is scattered into $2\pi$ steradians.
To measure the scatter, we scan a calibrated CCD camera in altitude (from roughly $15^\circ$ to $75^\circ$ 
angle of incidence) and record the number of counts registered.
By assuming the scattering is azimuthally symmetric, we compute the BRDF and subsequently the total integrated scatter (TIS) over the sampled altitude.
The result for one of our mirrors is shown in figure~\ref{fig:ScatterMeasurement}.
We found a wide variation in the level of scatter of our coatings, even after several rounds of cleaning.
The lowest TIS achieved was 2.7(5)\,ppm.
For two of the other mirrors, we measured TISs of 17.9(1.0)\,ppm, and 17.0(1.0)\,ppm.
The TIS of the fourth mirror was not measured after its final cleaning.

The scatter values measured by this method are slightly lower than, but within the errors of, the scatter values given in table~\ref{tab:CavityParams}, which are inferred from the total cavity losses.
However, our BRDF measurements did not measure the power scattered at small angles ($<15^\circ$), and will thus underestimate the total scatter.

\section{Discussion}
\label{sec:Discussion}

\subsection{Implications for gravitational wave detectors}

\begin{figure}[tbp]
    \centering
    \includegraphics[width=\columnwidth]{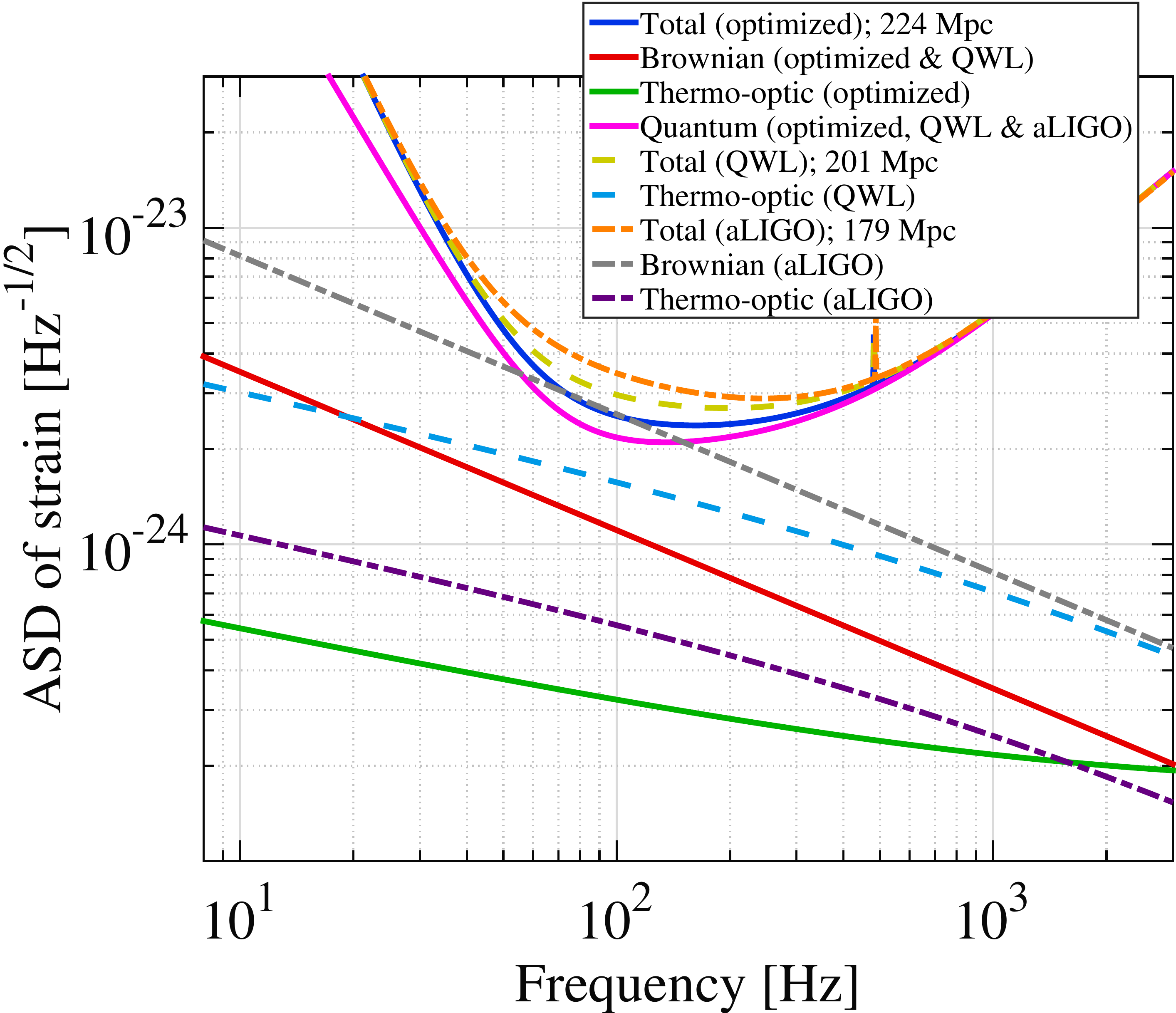}
    \caption{Strain curve for a gravitational wave detector with fused silica 
      test masses coated with optimized AlGaAs coatings.
    For comparison, the analogous strain curves for QWL AlGaAs coatings, and 
    for silica/tantala coatings, is shown.
    All other parameters are assumed identical to those of Advanced LIGO.}
    \label{fig:AlGaAsLIGO}
\end{figure}

Using the method described in section~\ref{sec:Optimization}, we have carried out a coating optimization for a version of Advanced 
LIGO in which the test mass coatings are fabricated from AlGaAs-based crystalline coatings and transferred to fused silica substrates.
The optimization was run with 29 layers on each input test mass and 81 layers on each end test mass.
In figure~\ref{fig:AlGaAsLIGO} we plot the resulting strain curve for such a detector, along with the strain curve for a version of Advanced LIGO with QWL AlGaAs coatings (28 layers on each input test mass, and 78 layers on each end test mass), and the strain curve for Advanced LIGO with as-built titania-doped silica/tantala coatings.
Both the optimized coatings and the QWL coatings produce a Brownian noise whose strain ASD is $\sqrt{S\mss{cBr}(f)} = (1.1\times10^{-23}\,\text{Hz}^{-1/2})\times(1\,\text{Hz}\ /\,f)^{1/2}$, which is a factor of 3 below the anticipated ASD for the currently installed silica/tantala coatings.
However, our optimized coatings offer superior thermo-optic noise performance across the entire gravitational wave band (10\,Hz to 7\,kHz), compared to the QWL coatings.
As a result, the detector's binary neutron star inspiral range~\cite{Sutton2013} increases from 201\,Mpc for the QWL coatings to 224\,Mpc for the optimized coatings. The anticipated range is 179\,Mpc for the silica/tantala coatings.

\subsection{Conclusion}
\label{sec:Conclusion}
We have demonstrated cancellation of photothermal noise in high-reflectivity substrate-transferred AlGaAs coatings.
This cancellation was achieved by optimizing the coating layer structure in such a way that the thermo-elastic and thermo-refractive contributions to the thermo-optic noise destructively interfere, hence minimizing the total coating thermo-optic noise.
Our result for the per-mirror absorption coefficient ($\alpha=0.05\,\text{cm}^{-1}$) is consistent with the result of Steinlechner et~al.~\cite{Steinlechner2015b}, who used photothermal common-path interferometry on a set of QWL AlGaAs high-reflectors to arrive at an absorption coefficient of $0.06\,\text{cm}^{-1}$.

The measured absorption and scatter from these coatings are promising, but improvement in the optical performance will be required for use in future gravitational wave interferometers.
For Advanced LIGO, requirements for scatter and absorption are set at the few-ppm and sub-ppm level, respectively~\cite{Harry2007}, and the requirements for third-generation interferometers may be even more stringent.
Additionally, for kilometer-scale interferometers, the spot sizes on the test masses can be more than 5\,cm.
This requires coating diameters which are tens of centimeters in diameter in order to avoid significant clipping losses of the beam.
In order for these coatings to be viable for gravitational wave detectors, the
substrate transfer process must be scaled up by more than a factor of 10 in diameter.

\begin{acknowledgments}
We gratefully acknowledge the informative discussions we have had regarding thermal noise with the Optics and Advanced Interferometer working groups of the LIGO Scientific Collaboration.
In particular, we appreciate the input from Yuri Levin, Matthew Evans, Johannes Eichholz, Andri Gretarsson, and Kazuhiro Yamamoto.
We thank Geoffrey Lovelace for pointing out an error in an earlier version of this manuscript.
TC, EDH, FS, KA, EKG, and RXA acknowledge support from the National Science Foundation under PHY-0757058.
GDC acknowledges support from EURAMET/EMRP (QESOCAS).
GDC and MA acknowledge support by the Austrian Science Fund (FWF) through project AI0090921.  
A portion of this work was performed in the UCSB Nanofabrication Facility.
RXA gratefully acknowledges funding provided by the Institute for Quantum Information and Matter, an NSF Physics Frontiers Center with support of the Gordon and Betty Moore Foundation.
JRS is supported by NSF award PHY-1255650.
This article has the internal project designation LIGO-P1500054.

\end{acknowledgments}

\appendix

\section{Coating structure}
\label{sec:CoatingStructure}

\begin{table}[tbp]
    \centering
    \begin{tabular}{l l l l l l}
        \toprule
        Layer  & \multicolumn{5}{c}{$d_{\!j}/\lambda_{\!j}$} \\
        \colrule
    \hphantom{0}1--5   & 0.1896 &  0.1121 &  0.4995 &  0.1000 &  0.4598 \\
    \hphantom{0}6--10  & 0.1695 &  0.2760 &  0.2145 &  0.2510 &  0.2388 \\
        11--15 & 0.2403 &  0.2508 &  0.2368 &  0.2553 &  0.2375 \\
        16--20 & 0.2571 &  0.2391 &  0.2564 &  0.2414 &  0.2550 \\
        21--25 & 0.2437 &  0.2533 &  0.2459 &  0.2515 &  0.2480 \\
        26--30 & 0.2498 &  0.2498 &  0.2482 &  0.2514 &  0.2469 \\
        31--35 & 0.2528 &  0.2457 &  0.2539 &  0.2447 &  0.2549 \\
        36--40 & 0.2439 &  0.2556 &  0.2433 &  0.2562 &  0.2427 \\
        41--45 & 0.2566 &  0.2423 &  0.2571 &  0.2420 &  0.2574 \\
        46--50 & 0.2417 &  0.2577 &  0.2414 &  0.2579 &  0.2412 \\
        51--55 & 0.2581 &  0.2409 &  0.2585 &  0.2405 &  0.2587 \\
        56--57 & 0.2401 &  0.2556 \\
        \botrule
    \end{tabular}
    \caption{Optical thickness of each coating layer, in fractions of a wavelength. 
      Odd-numbered layers are GaAs ($n_\text{H} = 3.51$), and even-numbered 
      layers are Al$_{0.92}$Ga$_{0.08}$As ($n_\text{L} = 3.00$).}
    \label{tab:CoatStruct}
\end{table}

In table \ref{tab:CoatStruct} we give the thickness of each layer of our AlGaAs coatings, in terms of optical thickness $d_j/\lambda_j$.
For each layer, the physical thickness $d_j$ can only be controlled to the nearest 50\,pm.
Additionally, there is systematic error in the thickness control for the GaAs and Al$_{0.92}$Ga$_{0.08}$As layers; the fractional uncertainties for this error are 0.5\,\% and 1.0\,\%, respectively.

\section{Thermo-optic material parameters}
\label{sec:TOparams}

The effective CTE of the coating is~\cite[eqs. A1--A3]{Evans2008}
    \begin{equation}
        \alpha_\text{c} = \sum_{j=1}^N \alpha_j\, \frac{d_j}{d}\, \frac{1+\sigma_\text{s}}{1-\sigma_j}
            \left[\frac{1+\sigma_j}{1+\sigma_\text{s}} + (1-2\sigma_\text{s})\frac{E_j}{E_\text{s}}\right],
    \end{equation}
the effective CTR of the coating is
    \begin{equation}
        \beta_\text{c} = -\frac{1}{\lambda_0} \frac{\partial x\mss{TR}}{\partial T}.
    \end{equation}
where $x\mss{TR}$ is the change in the sensed mirror position due to thermorefractive effects.
For a coating made entirely of QWL structure with high index material ($n_\text{H}$) as a top layer, $\beta_\text{c}$ can be approximated by~\cite[eq.~A.14]{Gorodetsky2008}
 \begin{equation}
        \beta_\text{c} \approx \frac{B_\text{H} + B_\text{L}}{4\bigl(n_\text{H}^2 - n_\text{L}^2\bigr)},
    \end{equation}
where $B_X$ is the fractional change in optical path length with respect to temperature in material X~\cite[eq.~31]{Yamamoto2013}\footnote{The value for $B_\text{X}$ here follows the calculation laid out by Yamamoto~\cite{Yamamoto2013}, which is slightly different from what is used by Evans et~al.~\cite[eqs.~A1, B8, B15]{Evans2008}. 
Since thermorefractive noise arises from the phase shift of the beam propagating back and forth inside the coating, the imaginary force used in the direct approach has to be applied on both sides of the coating. 
Thus, the deformation in the coating is not related to the substrate, and the effective thermal expansion is only corrected by the Poisson ratio of the coating.
All calculations and optimizations done in this work use this notion for $B_\text{X}$.}:
 \begin{equation}
        B_\text{X} = \beta_\text{X} + n_\text{X} \alpha_\text{X}\frac{1+\sigma_\text{X}}{1-\sigma_\text{X}}.
    \end{equation}
 
The effective heat capacity of the coating is~\cite[eq.~A4]{Evans2008}
    \begin{equation}
        C_\text{c} = \sum_{j=1}^N C_j \frac{d_j}{d}.
    \end{equation}

The effective thermal conductivity of the coating is
    \begin{equation}
        \kappa_\text{c} = \left(\sum_{j=1}^N \frac{1}{\kappa_j} \frac{d_j}{d}\right)^{-1}.
    \end{equation}

\bibliography{bib/algaasbib.bib}

\end{document}